\documentstyle[12pt,psfig,a4,cite]{article}
\begin{document}
\pagestyle{empty}
\begin{center}
{\bf A LOW-NUSSINOV MODEL FOR \Large $\gamma^* p\rightarrow V p$\\
}
~\\~\\
{\bf I. Royen}\\ 
{\small Inst. de Physique, U. de Li\`ege,
B\^at. B-5, \\
Sart Tilman, B4000 Li\`ege, Belgium\\}
\end{center}
\vskip 2cm
\begin{center}
{\bf Abstract}
\end{center}
\begin{quote}
\noindent We show that the lowest-order QCD calculation in a simple model 
of elastic vector-meson 
production does reproduce correctly the ratios of cross sections
for $\rho$, $\phi$ and $J/\psi$ (measured by H1$^{\citen{dataHp})}$ and ZEUS$^{\citen{dataZeus})}$), 
both in photoproduction and in high-$Q^2$ quasi-elastic scattering.
The mass and $Q^2$ dependence of the cross sections is reproduced as well. 
We also examine the lower-energy data, and see that the energy
dependence of the cross section does not depend on $Q^2$, and that a 
comparison with $\rho$ and $\phi$ data from NMC$^{\citen{datalowe})}$ suggests a soft pomeron intercept.
\end{quote}
\vskip 2cm
{W}e$^{\citen{JRCIR})}$ attempt to model Elastic Vector-meson production in the HERA kinematical range, i.e high energy and low t.
Several models$^{\citen{JRCIR,DL3,Ryskin})}$ have been proposed to describe this process. 
But so far, there has not been a model applied to the full range of masses and $Q^2$.  This is the object of our model$^{\citen{JRCIR})}$. 

The process that we study is the elastic interaction between a (virtual) photon and a proton: $\gamma^* p \rightarrow V p$.  
$V$ is the vector meson of mass $M_{V}$ and can be a $\rho$, $\phi$ or $J/\psi$.
The differential cross section of this process is given by:
${d\sigma_{\gamma p}/ dt}=
{|A|^2/(16\pi w^4)}$,
where $w^2=(P+q)^2$ is the centre of mass energy in the $\gamma^*p$ system.

Our model for exclusive vector-meson production includes three simple sub-models:
we use a non relativistic model for the vector meson, where the quark and the anti-quark share the $V$-momentum equally and have the same mass equal to $M_V \over {2}$, the colour-singlet
exchange is modeled \`a la Low-Nussinov (two perturbative gluon exchange), and we shall only
consider the constituent quarks of the proton.  This leads us to two form factors for the proton which enable us to cancel the infrared singularity that would result from the pole in the gluon propagator.  
In the following we call $\hat w^2\equiv (p+q)^2$, where $p$ is the parton four-momentum.  Moreover, we work in a Lorentz frame such that parton masses can be neglected.
The key point in the calculations is to get the leading term in the high-energy limit
 correctly, i.e $\hat w^2 \gg m_V^2$, $Q^2$ (where $Q^2=-q^2$ is the off-shellness of the photon).

In principle, 72 diagrams contribute to the amplitude.
However, with the use of those simplest sub-models and in the high-energy limit, the calculation of each part of the amplitude can be greatly simplified and we end up only with two diagrams shown in Fig.1.
\begin{figure}
\centerline{
\hbox{
\psfig{figure=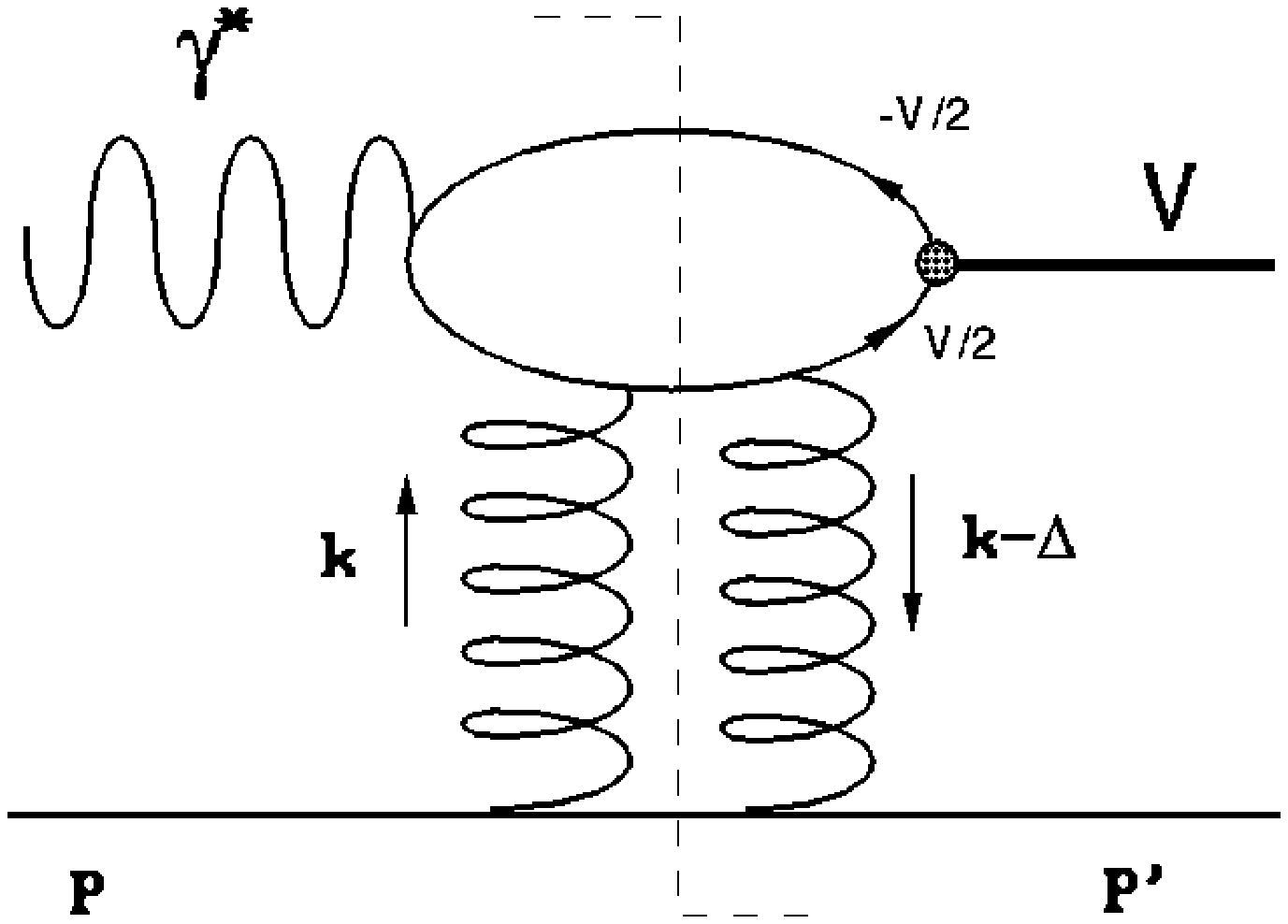,bbllx=2.5cm,bblly=15.5cm,bburx=17cm,bbury=26cm,width=5cm}
\hglue 1cm
\psfig{figure=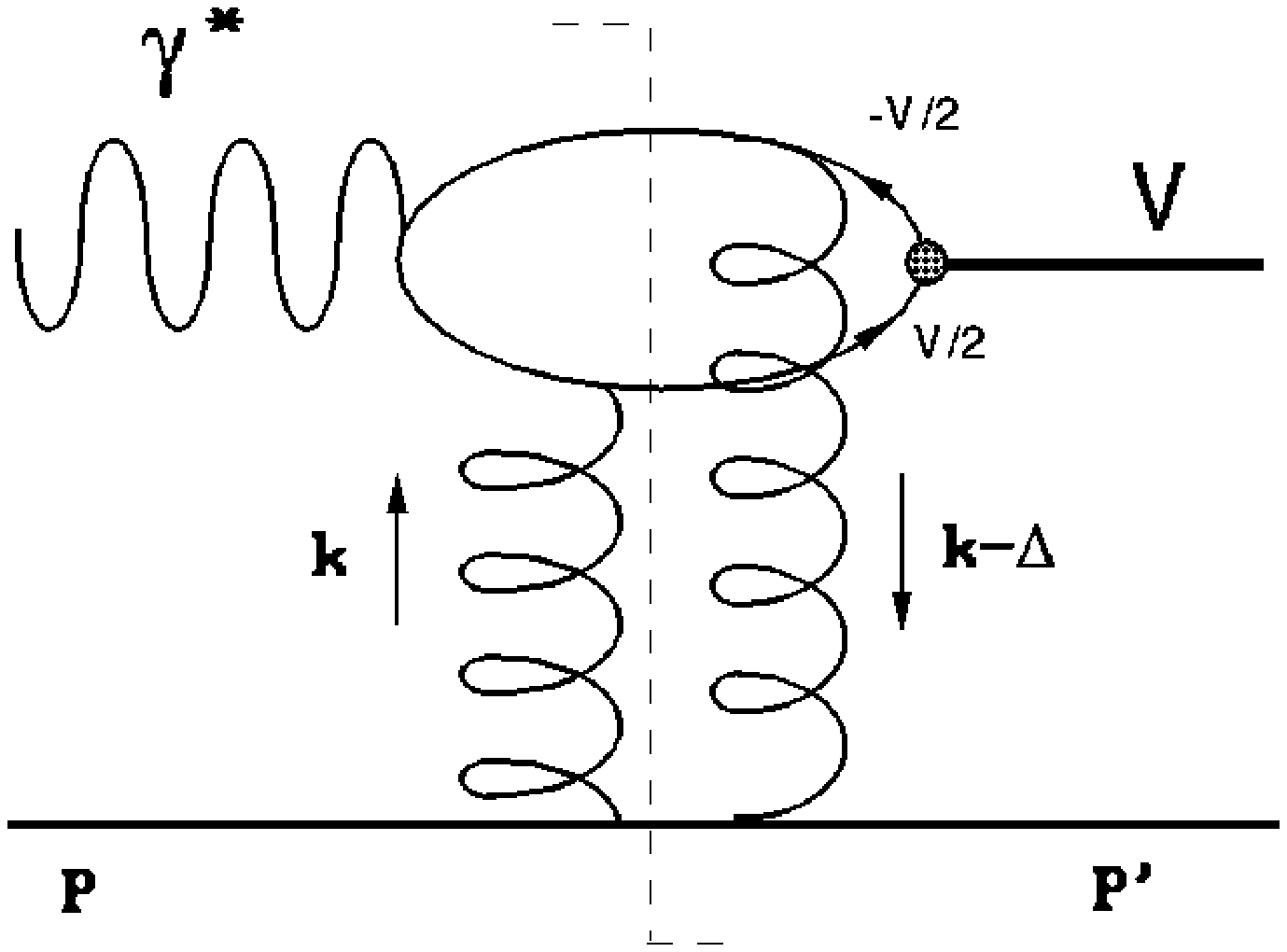,bbllx=2.5cm,bblly=15.5cm,bburx=17cm,bbury=26cm,width=5cm}}}
\begin{quote}
\baselineskip 0.7pt
{\small Figure~1:  The two diagrams accounting for the transition
$\gamma^*\rightarrow V$.
The dashed line represents the cut which puts the intermediate state on-shell.
}\end{quote}
\vglue -0.5cm
\end{figure}
It will come out that the imaginary part of the amplitude is proportional to $\hat w^2$.  
Crossing symmetry and analyticity then imply
that the amplitude is purely imaginary, up to terms of order $1/\hat w^2$, because the exchange is $C=+1$.  In order to calculate this imaginary part of the amplitude, we use Cutkovsky's rules, putting intermediate quarks on-shell and replacing their propagator by a delta function.

Now taking everything into account, we obtain the following expressions for the
amplitude, for the various possible helicities:
\begin{eqnarray}
{\cal A}(T\rightarrow T)&\propto&i\hat w^2{\Large R}\alpha_S^2
\nonumber\\
&&\hskip -2.3cm \times \int \frac{d^2k}{k^2(k-\Delta)^2}
{[{\cal E}_1(t)-{\cal E}_2(k,k-\Delta)]
\ [k\cdot(k-\Delta)]\over \left(t - m_V^2 - Q^2 + 4\ k\cdot(k-\Delta) \right)\
(m_V^2 + Q^2 - t)}\\
{\cal A}(L\rightarrow L)&=&{\sqrt{Q^2}\over m_V}\times {\cal A}(T\rightarrow
T)
\end{eqnarray}
The first $T$ ($L$) refers to the polarisation of the photon and the second to that of the meson.
As previously advertised, this answer is proportional to $\hat w^2$, and is thus purely imaginary.
One also finds that the longitudinal amplitude is proportional to $Q$ times the 
transverse amplitude and that the helicity violating amplitudes ${\cal A}(L\rightarrow T)$, ${
\cal A}(T\rightarrow L)$ are nonleading in energy, i.e. down by $1/\hat w^2$.  

The resulting differential cross section is given by:
\begin{equation}
{d\sigma\over dt}={d\sigma_T\over dt}+\varepsilon {d\sigma_L\over dt}
={R\over 16\pi \hat w^2} \left[|A(T\rightarrow T )|^2+\varepsilon|A(L\rightarrow L
)|^2\right]
\end{equation}
with $\varepsilon$ the polarisation of the photon beam: $\varepsilon\approx 1$
at HERA
and $\varepsilon\approx 0.75$ at NMC.
Clearly this model cannot say anything about the energy dependence
of the cross section. We shall assume that it comes in as a factor, $R$ (Regge factor),
and check whether the latter is mass- or $Q^2$-dependent. 
\vskip 0.3 cm
We first give the results that we obtain
for the various cross sections measured by
ZEUS and H1.  In Fig.~2 we show the dependence on $Q^2$ and
$m_V$ of the integrated elastic cross section $\sigma(Q^2)$.
We see that a common (Regge) factor is consistent with the data
taken at HERA, as shown in Fig.~2(a). We insist on the fact that this factor
is independent both of $Q^2$ and of $m_V$, as one would expect within
Regge theory.
Although we see no reason why our model should work in photoproduction,
it turns out that our curves do go
through the photoproduction points.
The result of a fit  
$\sigma\propto (Q^2)^{-n}$ at large $Q^2$, gives $n_{\rho,\phi}=2.5$ and $n_{J/\psi}=1.5$
which is in good agreement with experiments.
\begin{figure}
\centerline{
\psfig{figure=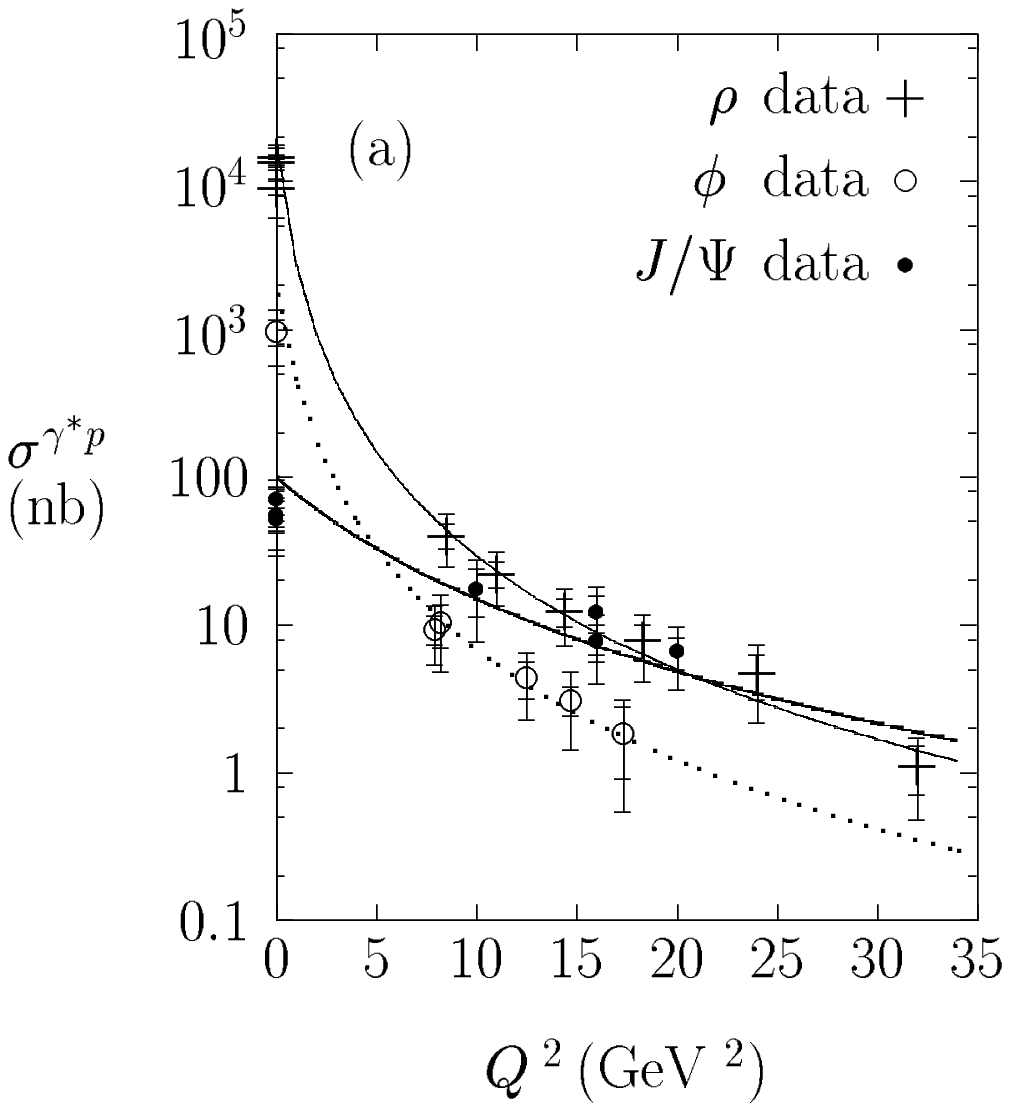,height=7.5cm}\ \ \ \ \
\psfig{figure=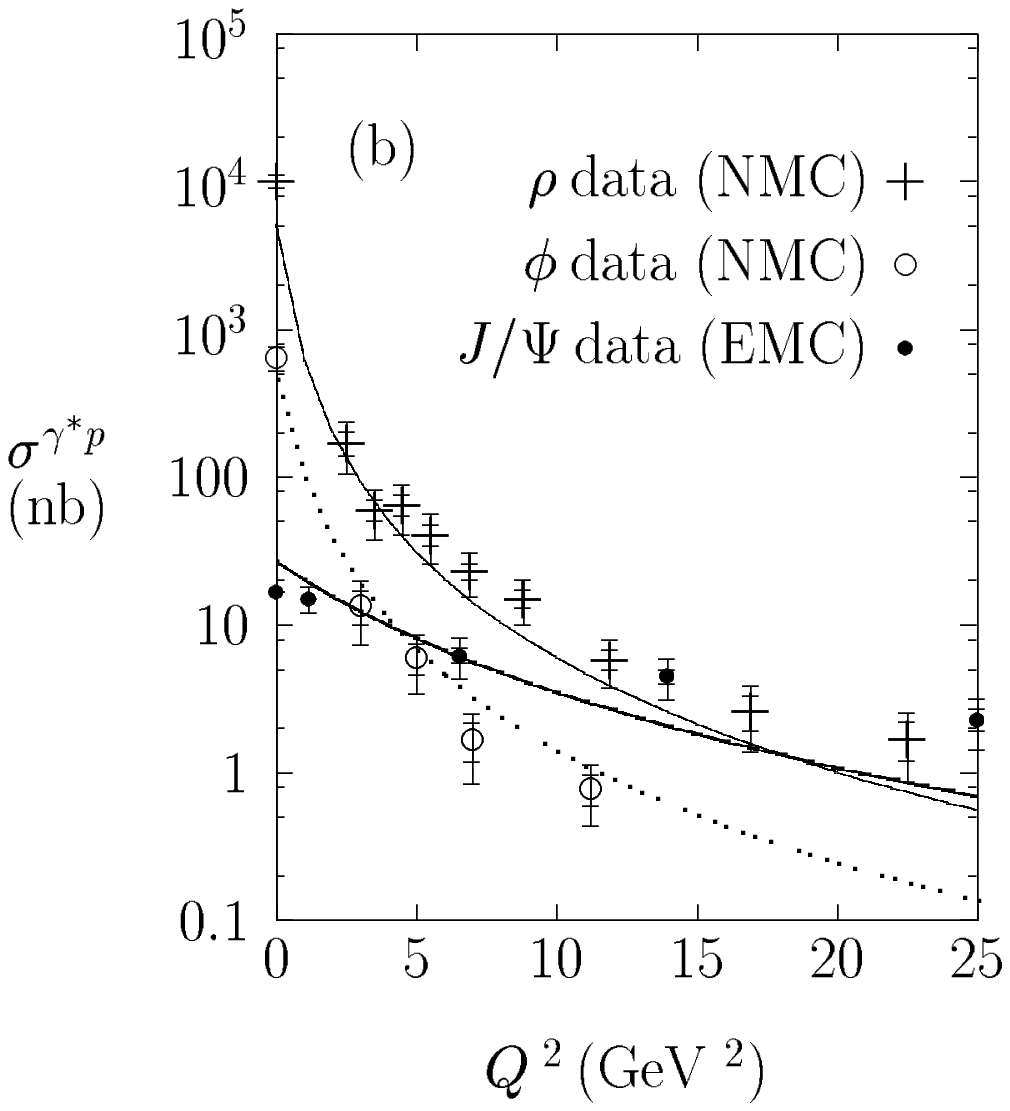,height=7.5cm}}
\begin{quote}
\baselineskip 0.7pt
{\small Figure~2: (a) Cross sections as functions of $Q^2$, compared
with data from H1$^{~\citen{dataHp})}$ and ZEUS$^{~\citen{dataZeus})}$
at $<w>\approx$100~GeV; (b)  Cross sections compared to lower-energy EMC
and NMC data$^{~\citen{datalowe})}$, at \hbox{$<w^2>\approx$200~GeV$^2$}.
}\end{quote}
\vskip -0.4 cm
\end{figure}
\begin{figure}
\centerline{
\psfig{figure=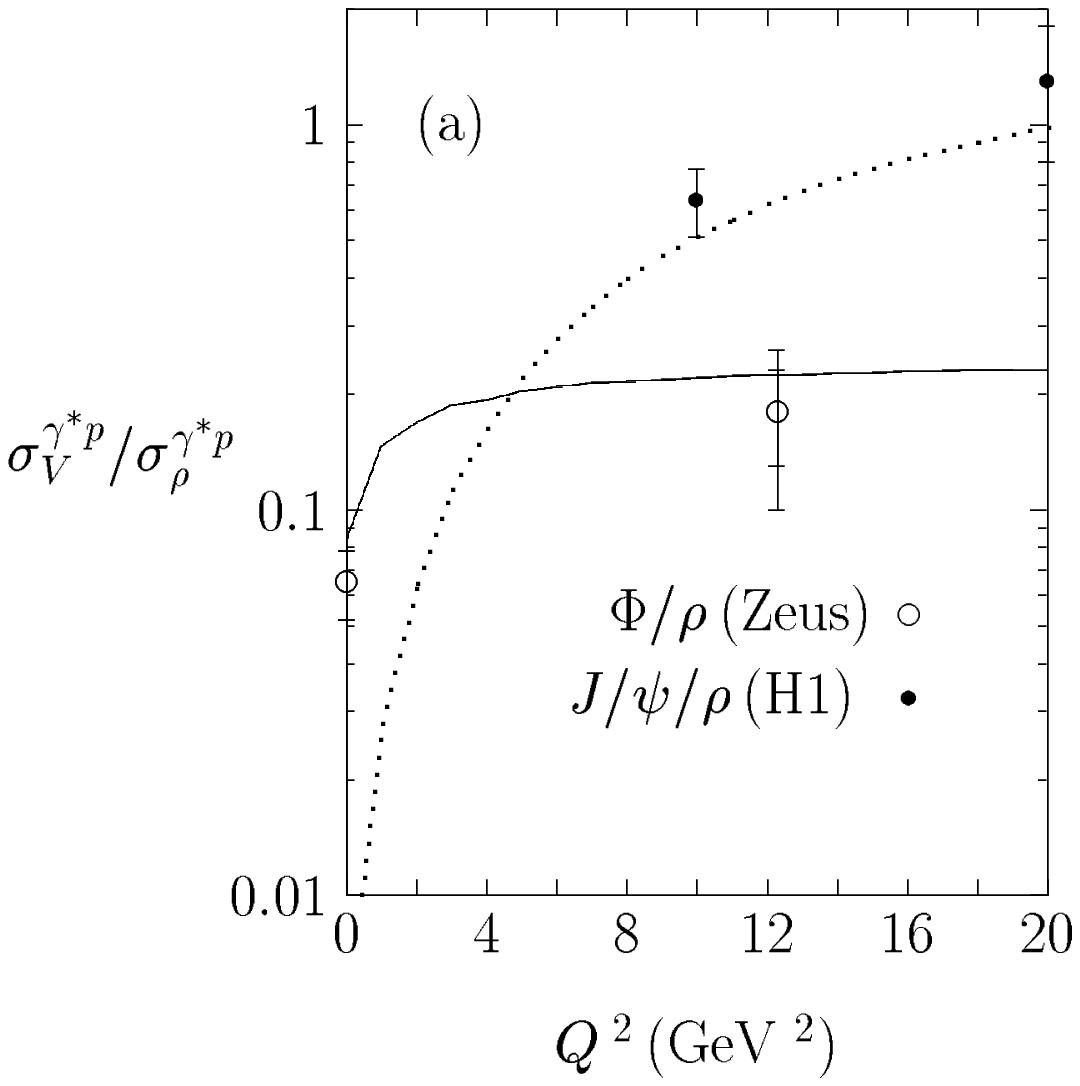,height=7.5cm}\ \ \ \
\psfig{figure=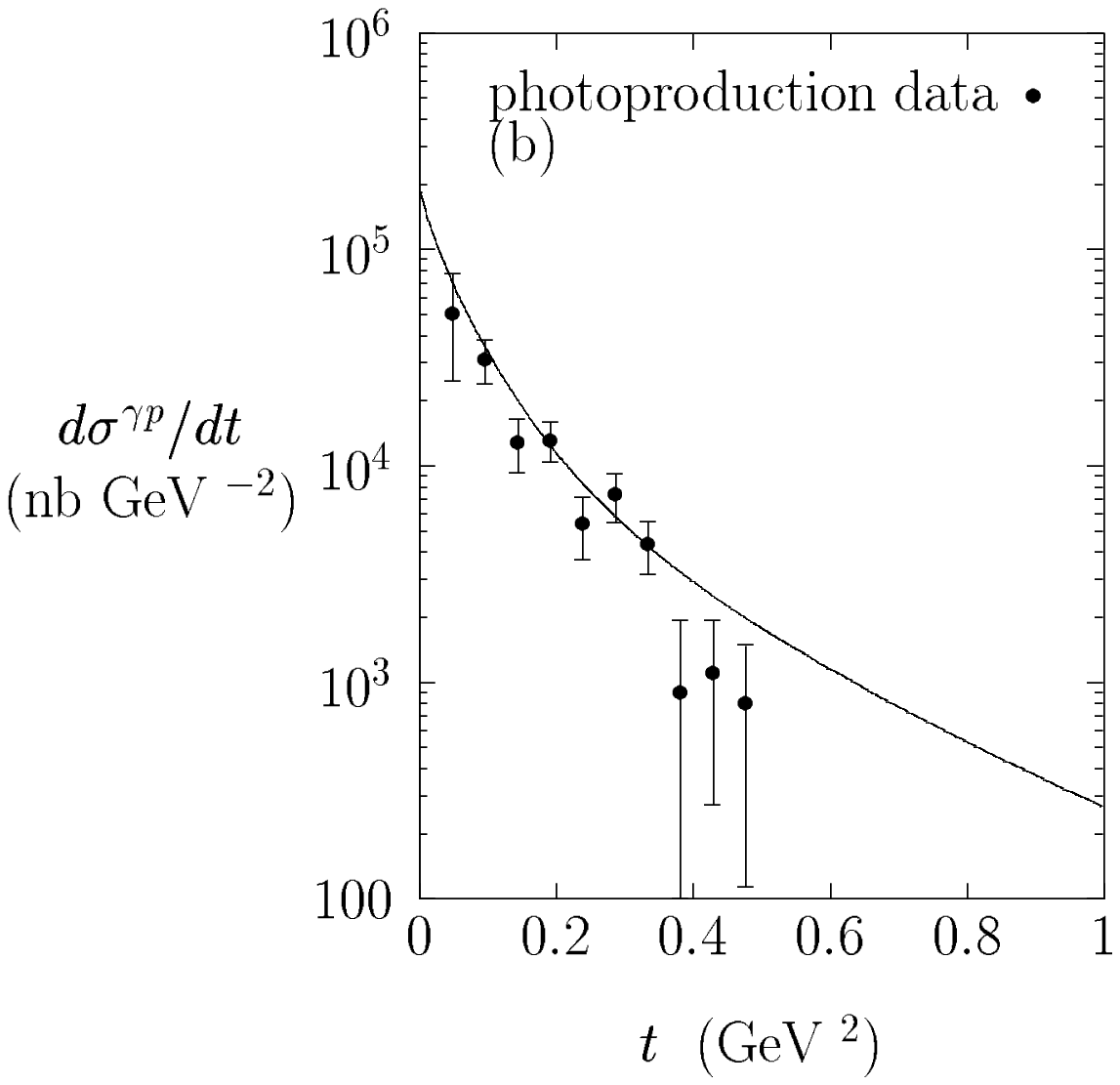,height=7.5cm}}
\begin{quote}
\baselineskip 0.7pt
{\small Figure~3: (a) Ratio of cross sections as functions of $Q^2$
at \hbox{$<w>\approx$100~GeV}, compared with data from H1$^{\citen{dataHp})}$
and Zeus
$^{\citen{dataZeus})}$;
(b) Photoproduction differential cross section
$d\sigma/dt$, compared with H1 data$^{\citen{dataHp})}$.
}\end{quote}
\vskip -0.4 cm
\end{figure}

In the ratio of cross sections, some of the systematic uncertainties
cancel, and the reproduction of that ratio is a more stringent test of
our model, especially as the normalisation then drops out of our prediction.
We show in Fig.~3(a) the result of such a comparison.
Again, we see that our model fares well, even in photoproduction.

Other results are for the differential cross section.
To illustrate the effect of the curvature, we compare in Fig.~3(b) our results
with the data for $\rho$ photoproduction in H1. We see that, here again, our curve reproduces the data fairly well, and indicates that 
$d\sigma/dt$ is {\it not} an exponential.

Finally, we can now examine the $w^2$-dependence
of the cross sections, and compare with NMC and EMC data. 
We have seen that at HERA, the Regge factor does not 
depend either on the meson mass or on $Q^2$.
We adopted the same philosophy when fitting to lower-energy cross sections from
EMC and NMC$^{\citen{datalowe})}$,
and we observed that $\rho$ data point to a soft pomeron intercept.  There may be a problem with the smallness of the $\phi$ cross section at NMC which implies a larger intercept.  We do not believe that much can be concluded from the $J/\psi$ data, which seem to have an inelastic background.
We show in Fig.~2(b) the curves which correspond to an intercept of 1.16.
We see that the fit is reasonable for the $\rho$ and the $\phi$.
We can see that there is no sign in either $\rho$ or $\phi$ data of a $Q^2$
dependence of the intercept.

To sum up, both the $m_V$ and the $Q^2$-dependence of the cross sections, 
as well as the $t$-dependence can be understood in a simple QCD model.  
The HERA data indicate that the Regge factor does not depend either 
on $Q^2$ or $m_V$.  The comparison with NMC $\rho$ data points to a soft 
pomeron intercept.  The only problem we have is the prediction of 
the ratio $\sigma_L/\sigma_T$, which may be resolved by the introduction of Fermi momentum.

\goodbreak
\end{document}